\documentclass[prl,twocolumn,showpacs,amsmath,amssymb]{revtex4}

\usepackage{graphicx}
\usepackage{dcolumn}
\usepackage{bm}
\usepackage{epsfig}
\usepackage{amsmath}
\usepackage{color}

\newcommand{\gam} {\gamma}
\newcommand{\kap} {\kappa}

\newcommand{\aop} {a}

\newcommand{\caop} {a^{\dag}}

\newcommand{\beq} {\begin{equation}}
\newcommand{\eeq} {\end{equation}}
\newcommand{\bseq} {\begin{subequations}}
\newcommand{\eseq} {\end{subequations}}
\newcommand{\beqz} {\setlength{\mathindent}{0cm}\begin{equation}}
\newcommand{\eeqz} {\end{equation}}
\newcommand{\ber} {\begin{eqnarray}}
\newcommand{\eer} {\end{eqnarray}}
\newcommand{\bers} {\begin{eqnarray*}}
\newcommand{\eers} {\end{eqnarray*}}

\begin{document}

\title{Two-photon gateway in one-atom cavity quantum electrodynamics}

\author{A. Kubanek}
\author{A. Ourjoumtsev}
\author{I. Schuster}
\author{M. Koch}
\author{P.W.H. Pinkse}
\author{K. Murr}
\author{G. Rempe}
\affiliation{Max-Planck-Institut f\"ur Quantenoptik,
Hans-Kopfermann-Str.~1, D-85748 Garching, Germany}
\date{\today}

\begin{abstract}
Single atoms absorb and emit light from a resonant laser beam photon by photon. We show that a single atom strongly coupled to an optical cavity can absorb and emit resonant photons in pairs. The effect is observed in a photon correlation experiment on the light transmitted through the cavity. We find that the atom-cavity system transforms a random stream of input photons into a correlated stream of output photons, thereby acting as a two-photon gateway. The phenomenon has its origin in the quantum anharmonicity of the energy structure of the atom-cavity system. Future applications could include the controlled interaction of two photons by means of one atom.
\end{abstract}

\maketitle
Atom-light interactions at the single-particle level have always been a central theme in quantum optics. A cornerstone of this research is the study of the photon statistics of the light resonantly scattered by a single atom. Photon antibunching, i.e. the sequential emission of single photons, is by now a well established phenomenon, confirming Einstein's view that energy is radiated quantum by quantum. The situation, however, is different if the atom is forced to emit and absorb the photon several times, as is possible if the atom is placed between cavity mirrors. In this case the combined atom-cavity system becomes the light source under investigation. In fact, for strong coupling between light and matter novel photon statistics have been predicted and observed for many intracavity atoms \cite{Rempe91,Foster00,Foster00a}. For one atom in the cavity, photon antibunching has been demonstrated \cite{Kuhn02,McKeever04,Keller04,Birnbaum05,Hijlkema07,Dayan08}.

Here we address the question whether a single atom can simultaneously absorb and emit two resonant photons. Such an effect could allow interactions between two photons mediated by one atom, with interesting applications including a single-atom single-photon transistor \cite{Chang07}. Towards this goal we place the atom inside a high-finesse optical cavity, operated in the strong-coupling regime, and tune the system into the nonlinear regime of cavity quantum electrodynamics \cite{Schuster08}. Specifically, with a laser resonant with the atom, we selectively populate in a two-photon process a quantum state of the combined atom-cavity system containing two energy quanta. The decay of this state then leads to the correlated emission of two photons. Conversely, the corresponding photon bunching has been proposed \cite{Carmichael96,Schneebeli08} as a means to detect the so-called higher excited Jaynes-Cummings states \cite{Jaynes63}. These states are at the heart of considerable experimental efforts which go far beyond the atomic physics community \cite{Mabuchi02,Vahala03,Khitrova06,Faraon08}, owing to their remarkable properties regarding atom-field entanglement and from the general perspective that they represent an elementary structure of a fermion-boson system. In the microwave domain, they have been featured in numerous publications for several decades \cite{Rempe87,Brune96,Schuster07,Fink08,Hofheinz08}. In the optical domain, however, they have escaped an experimental observation only until recently \cite{Schuster08}. It is the optical domain with its availability of photon counting devices where these states fully unfold their unique potential in generating definite multi-photon states in a deterministic process.

\begin{figure}
\includegraphics[width=8cm]{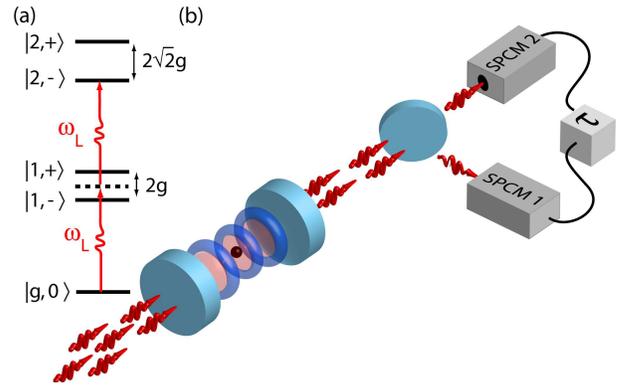}
\caption{\label{Figure1} The energy level structure of one atom strongly coupled to a quantized field, (a), governs the statistics of photons which leak out of the cavity, (b). Photons arrive randomly at the input mirror and exit in pairs as soon as two laser photons are on resonance with the two-photon dressed state $|2,\pm\rangle$.}
\end{figure}
When the electromagnetic interaction between a single atom and the light field is strong enough, the atom-light system exhibits a completely new structure, different from the sum of its parts. For a two-state atom coupled to an optical mode (between two mirrors) with successive photon number states, $|0\rangle,|1\rangle,|2\rangle ...$, the dressed states \cite{Jaynes63} consist of a ground state and a discrete ladder of pairs of states, $|1,\mp\rangle,|2,\mp\rangle, ...$, see Fig.\ref{Figure1}a. The strategy is to send photons onto the input mirror, with each photon being resonant with the atomic energy (here represented by a dashed line in Fig.\ref{Figure1}a). In this case, and for a suitable cavity frequency, the two-photon doublet $|2,\mp\rangle$ is directly excited whereas the single-photon doublet $|1,\mp\rangle$ is completely avoided because the energy level structure is strongly anharmonic. This excitation entangles the atom with two photons and leads to an enhancement of photon pairs leaving the cavity through the output mirror, Fig.\ref{Figure1}b.

Photon pair emission can be revealed by intensity autocorrelations, traditionally quantified by the second-order normalized correlation function $g^{(2)}(\tau)$, usually measured with two single photon counters in a Hanbury Brown and Twiss configuration and defined as the ratio between the rate of clicks separated by a time delay $\tau$ and the rate of clicks separated by long time delays $|\tau|\rightarrow\infty$. For zero time delay, $\tau=0$, its expression in terms of the cavity mode creation and annihilation operators is $g^{(2)}(0)=\langle a^{\dagger2}a^2\rangle/\langle\caop\aop\rangle^2$. When the system is excited with a weak laser beam impinging on the input mirror of the cavity, the $g^{(2)}$ function of the transmitted light has been shown to be independent of the laser intensity \cite{Carmichael91,Brecha99,Goto04}.

\begin{figure}
\includegraphics[width=8cm,height=6cm]{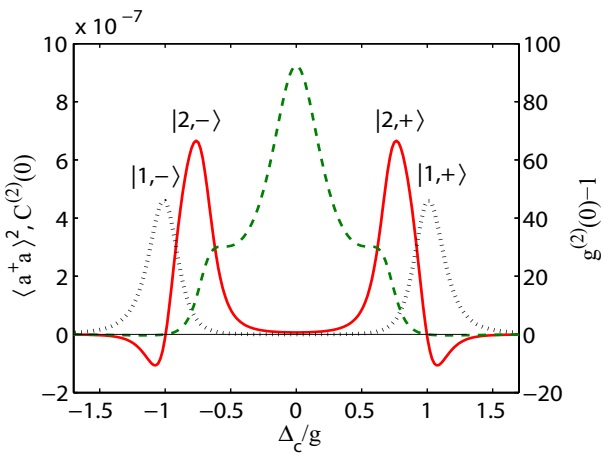}
\caption{\label{Figure2} Photon number squared (dotted line) and two-photon correlation functions versus the cavity detuning with an input field corresponding to 0.01 photon in an empty cavity. The normalized $g^{(2)}(0)-1$ function (dashed line) presents a maximum at zero cavity detuning, $\Delta_c=0$, so that the two-photon dressed states $|2,\mp\rangle$ appear as shoulders. In contrast, the differential correlation function $C^{(2)}(0)$ (solid line) has clear maxima on the second dressed states. Notice that $C^{(2)}(0)$
is the product of the dotted and dashed lines. The negative
values for $C^{(2)}(0)$ and $g^{(2)}(0)-1$ on the dressed states $|1,\mp\rangle$ indicate sub-Poissonian emission of single photons (though hardly visible in this plot for $g^{(2)}(0)-1$).}
\end{figure}
The $g^{(2)}$ function could allow one to localize the multiphotonic higher-order states, as these should present strong photon-photon correlations. However, as we show below, a more appropriate choice for our purpose is the differential correlation function, $C^{(2)}(\tau)$, which at $\tau=0$ reads:
\beq
C^{(2)}(0)=\langle a^{\dagger2}a^2\rangle -\langle\caop\aop\rangle^2\ ,
\eeq
and which scales as the square of the input intensity at weak input fields.  Notice that $C^{(2)}(0)=[\,g^{(2)}(0)-1\,]\langle\caop\aop\rangle^2$. For a coherent intracavity field, one has $C^{(2)}(0)=0$, alike $g^{(2)}(0)=1$. Maximally sub-Poissonian light, $g^{(2)}(0)=0$, would correspond to the minimum negative value $C^{(2)}(0)=-\langle\caop\aop\rangle^2$, and super-Poissonian emission, $g^{(2)}(0)>1$, corresponds to $C^{(2)}(0)>0$.

The correlation function $C^{(2)}$ is less sensitive to single-photon excitations than $g^{(2)}$ and provides a clearer measure of the probability to create two photons at once in the cavity. To illustrate this point, we plot in Fig.~\ref{Figure2} the behavior of the system as a function of the detuning $\Delta_c=\omega_L-\omega_{cav}$ between the laser and the cavity mode. We assume the atomic frequency $\omega_a$ to be equal to the cavity frequency, i.e. $\omega_a=\omega_{cav}$, and that the atomic dipole and cavity field decay rates, $\gam$ and $\kap$ are small enough to be in the strong atom-cavity coupling regime, $g\gg(\gam,\kap)$ (the parameters for Fig.~\ref{Figure2} are $(\gam,g)=(3\kap,10\kap)$). As a result, the mean photon number squared (dotted line) shows two symmetric narrow peaks at the frequency of the normal modes, $|1,\mp\rangle$, whereas the two-photon states $|2,\mp\rangle$ do not contribute to the photon number for these frequency parameters and weak input intensity. In this regime, the mean photon number squared gives the probability of preparing two single photons independently, $\langle\caop\aop\rangle^2=[P(g,1)]^2$, where $P(g,1)$ is the probability of having one photon in the cavity and the atom in its internal ground state $|g\rangle$. For the same parameters, the normalized correlation function $g^{(2)}(0)$ (dashed line) presents shoulders near the frequencies of the second dressed states, where the probability $P(g,2)$ to be in state $|g,2\rangle$ is maximized. However, there is a much higher maximum at the center, $\Delta_c=0$, precisely where the occupation probability $P(g,2)$ has a minimum. This happens because for $\Delta_c=0$ the probability of having uncorrelated photons is also small and, in fact, much smaller than $P(g,2)$. The height of this central peak\cite{Faraon08} dominates the frequency dependence of $g^{(2)}$, and could overlap with the second dressed-state resonances, which can be washed out by extra broadening mechanisms.

The situation is more favorable when using the differential correlation function (solid line), which at weak fields reads $C^{(2)}(0)=2P(g,2)-[P(g,1)]^2$. The maxima appear clearly at the detunings $\Delta_c=\mp g/\sqrt{2}$ of the second dressed states $|2,\mp\rangle$, owing to $2P(g,2)\gg [P(g,1)]^2$, whereas $C^{(2)}(0)$ has a minimum for $\Delta_c=0$ because it becomes the difference of small probabilities. Away from these resonances, one finds two minima on the normal modes $|1,\mp\rangle$ where the negative values of $C^{(2)}$ correspond to sub-Poissonian emission.

In the experiment, we use a high-finesse optical cavity that supports a TEM${}_{00}$ mode near-resonant with the $5^2 S_{1/2} F=3,m_F=3\rightarrow 5^2 P_{3/2} F=4,m_F=4$ transition of ${}^{85}$Rb atoms at wavelength $\lambda=780.2$ nm. The atomic polarization and cavity field decay rates are $(\gam,\kap)/2\pi=(3,1.3)$ MHz. This cavity mode is excited by near-resonant light impinging on one mirror, with the twofold purpose of probing the system as well as cooling the atom. The atoms injected into the cavity with an atomic fountain are caught by two superimposed optical dipole traps. The first trap is created with a far red-detuned laser ($785.3$ nm) resonantly exciting a two free-spectral ranges (FSR) detuned TEM${}_{00}$ mode supported by the cavity. The second trap ($775.2$ nm) is a sum of a TEM${}_{10}$ and a TEM${}_{01}$ mode, both two FSR blue-detuned with respect to the probe light \cite{Puppe07}. The resulting doughnut-shaped mode repels the atoms towards the cavity axis, thereby favoring events where atoms are strongly coupled to the cavity and decreasing the losses of the atoms in the radial direction. The initial atom-cavity detuning together with the induced Stark shift sets an effective atom-cavity detuning of $\approx-2\pi\times8.5$ MHz.

\begin{figure}
\includegraphics[width=8cm]{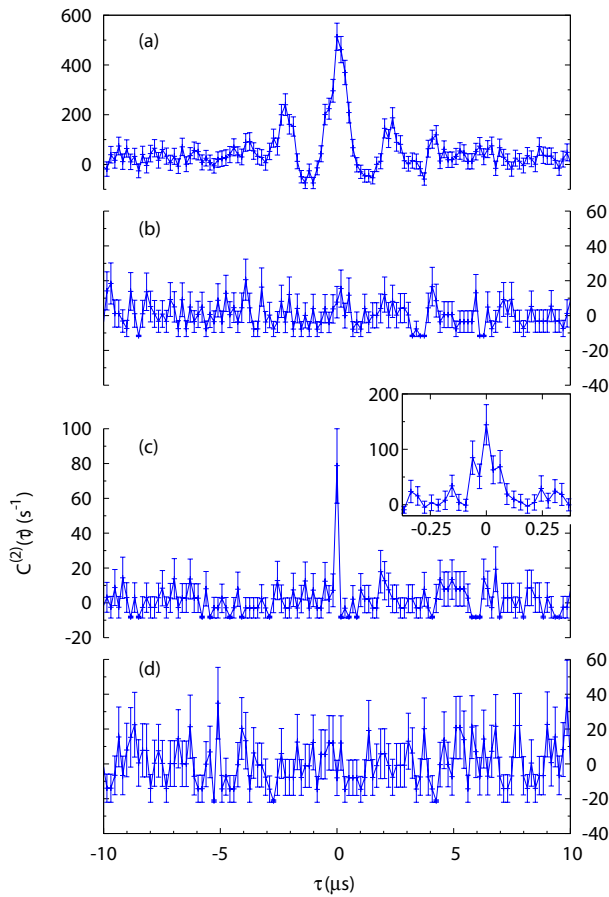}
\caption{\label{Figure3}
Time-dependent correlation function for different cavity detunings. The error bars are standard deviations. The detunings are for (a)-(d) $\Delta_c/2\pi=(0,-3,-10,-18)$ MHz. (a) shows large bunching with long-tail oscillations, yielding information on the micromotion of the atom in the trap. From (b)-(d) the correlations are produced by the atom-cavity system, notably showing a bunching in (c, and inset) at the two-photon resonance.}
\end{figure}

The cooling and trapping protocol as well as the selection of good coupling events has been described elsewhere \cite{Schuster08}. In brief, trapped atoms undergo a sequence of $500\ \mu s$ cooling periods, alternated with $100\ \mu s$ probing intervals. The intensity transmitted during these cooling periods allows us to determine the effective atom-cavity coupling constant $g$, and to postselect the events where this constant was sufficiently high ($g/2\pi\approx 11.5$ MHz). For the measurement, about $20000$ atoms were trapped in $127$ hours of pure measurement time. Each trapped atom starts a measurement sequence including $31$ probing intervals. About $7 \%$  of these intervals survived the selection procedure, which gives an effective probing time of $4$ seconds. The experimental correlation production rate was mainly limited by the atomic storage times and by the overall photon detection efficiency $(\sim5 \%$).

We determined $C^{(2)}$ by counting the number of photon clicks on the detector SPCM2 at time $t+\tau$ within a time window $\Delta\tau$, knowing that a photon has been detected at the detector SPCM1 at time $t$, and we subtract the averaged coincidence counts obtained for very long time delays ($\tau\gg 10 \kappa^{-1}$) when the photons are uncorrelated.
The dimensionless theoretical $C^{(2)}$, integrated within the time window $\Delta\tau$, is then compared to this experimental coincidence count rate after accounting for mirror transmission, losses and detection efficiency.

\begin{figure}
\includegraphics[width=8cm]{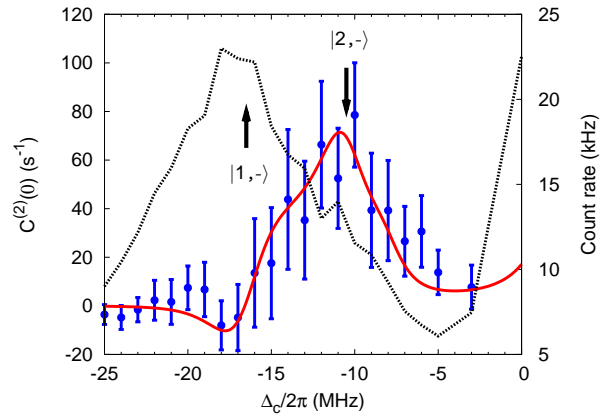}
\caption{\label{Figure4} Correlation spectrum for zero time delay as a function of the cavity detuning. The spectrum shows that the rate of coincidences (left scale) is maximum when two laser photons become resonant with the two-photon dressed state $|2,-\rangle$. The solid curve is from quantum theory which describes the interaction of one atom and two resonant photons (see text for details). Also shown is the measured photon count rate which is maximum on the single photon dressed state $|1,-\rangle$ (dotted, right scale); the standard deviations (0.2-1 kHz) are not shown for clarity.}
\end{figure}
Fig.~\ref{Figure3} shows $C^{(2)}$ as a function of the delay time $\tau$ and for different detunings. The size of the coincidence window is set to $\Delta\tau=170$ ns $\lesssim 2\kap^{-1}$. On the cavity resonance, $\Delta_c=0$, Fig.~\ref{Figure3}a, the expected photon statistics are completely dominated by the effect of the atomic motion, where we observe a large bunching with long-period oscillations at the characteristic axial trapping period ($2.2\ \mu s$)\cite{muenstermann99}. In this case, the micro-oscillations of the atom in the intracavity trap induce small variations in the coupling $g$, which in turn induce large fluctuations of the emitted light at $\Delta_c=0$. This phenomenon rapidly disappears when the probe frequency is detuned with respect to the cavity frequency, because in this case small variations in the coupling have little effect on the emitted light. This is already largely the case at $\Delta_c/2\pi=-3$ MHz, Fig.~\ref{Figure3}b, where we observe no oscillations for time scales above $\kap^{-1}$. Here, we find small values of $C^{(2)}(0)$, which is expected from quantum theory as one is away from any resonance of the coupled atom-cavity system. As we sweep the laser frequency further away from the cavity, however, we find bunching and super-Poissonian statistics, Fig.~\ref{Figure3}c, precisely at the two-photon resonance $|2,-\rangle$. In the inset of Fig.~\ref{Figure3}c we plot data that are gathered with a much higher time resolution, $\Delta\tau=30$ns, representing the sum of all the coincidences recorded for detunings around the second dressed state $|2,-\rangle$ within a range $\pm2\pi\times4$ MHz (see Fig.\ref{Figure4} for the detuning dependence of $C^{(2)}(0)$). It shows that two photons emitted by the atom-cavity system are correlated within a time $T_{corr}\lesssim150$ ns, with a HWHM ($\approx30-60$ ns) compatible with the lifetime $\Gamma_{-}^{-1}\approx33$ ns of state $|1,-\rangle$.
Eventually, Fig.~\ref{Figure3}d are data registered when we excite the dressed state $|1,-\rangle$ on resonance, and we see that the photons are now essentially uncorrelated. This is also consistent with theory, which predicts an antibunching \cite{Birnbaum05,Dayan08} too small to be observed with our cavity parameters (also see below).

Even though classical fields can produce photon bunching, the frequency dependence of the photon correlations should show resonances at the two-photon dressed states, which has no classical analogue \cite{Carmichael96}.
We have consequently sampled the spectrum every MHz across the normal mode $|1,-\rangle$ and across the two-photon dressed state $|2,-\rangle$. Fig.~\ref{Figure4} shows $C^{(2)}(0)$ as a function of the cavity detuning.
We observe a resonance over a frequency range $\approx2\pi\times6$ MHz with a peak center on the two-photon state.
The solid curve is obtained from fixed-atom theory for our parameters and shows an overall satisfactory agreement. We have cross-checked that for our parameters the numerical solution from the master equation including photon numbers higher than two is essentially the same as the analytical theory in the weak-field limit \cite{Carmichael91,Brecha99}; we can safely assume that three photon events (and higher) are negligible. The coupled atom-cavity system serves as a two-photon gateway which favors the transmission of twin photons through the cavity. The number of coincidences is enhanced in this region, with a coincidence count rate of more than $80$ per second of probing time, more than ten times larger compared to a coherent field of the same intensity ($g^{(2)}(0)\gtrsim 10$).

A remarkable feature of the atom-cavity system is its ability to react differently depending on whether it is excited by single photons or twin photons. This is striking when comparing the rate of coincidences to the photon count rate (dotted data). Here, we clearly see that the two-photon count rate is higher on the second dressed state $|2,-\rangle$ than on the first one, whereas the photon count rate is highest on the state $|1,-\rangle$. This asymmetry is a deep manifestation of the anharmonicity of the system owing to its discrete multiphotonic nature, here viewed in correlation spectroscopy.
With larger atom-cavity couplings, it should open new perspectives for using single atoms in controlled photonic quantum gates.

\begin{acknowledgments}
We warmly thank L. Orozco for fruitful discussions.
Partial support by the Bavarian PhD programme of excellence QCCC, the DFG research unit 635, the DFG cluster of excellence MAP and the EU project SCALA are gratefully acknowledged.
\end{acknowledgments}

\begin{thebibliography}{28}
\expandafter\ifx\csname natexlab\endcsname\relax\def\natexlab#1{#1}\fi
\expandafter\ifx\csname bibnamefont\endcsname\relax
  \def\bibnamefont#1{#1}\fi
\expandafter\ifx\csname bibfnamefont\endcsname\relax
  \def\bibfnamefont#1{#1}\fi
\expandafter\ifx\csname citenamefont\endcsname\relax
  \def\citenamefont#1{#1}\fi
\expandafter\ifx\csname url\endcsname\relax
  \def\url#1{\texttt{#1}}\fi
\expandafter\ifx\csname urlprefix\endcsname\relax\def\urlprefix{URL }\fi
\providecommand{\bibinfo}[2]{#2}
\providecommand{\eprint}[2][]{\url{#2}}

\bibitem[{\citenamefont{{Rempe} et~al.}(1991)\citenamefont{{Rempe}, {Thompson},
  {Brecha}, {Lee}, and {Kimble}}}]{Rempe91}
\bibinfo{author}{\bibfnamefont{G.}~\bibnamefont{{Rempe}} \textit{et~al.}},
  \bibinfo{journal}{Phys. Rev. Lett.} \textbf{\bibinfo{volume}{67}},
  \bibinfo{pages}{1727} (\bibinfo{year}{1991}).

\bibitem[{\citenamefont{{Foster}
  et~al.}(2000{\natexlab{a}})\citenamefont{{Foster}, {Orozco},
  {Castro-Beltran}, and {Carmichael}}}]{Foster00}
\bibinfo{author}{\bibfnamefont{G.~T.} \bibnamefont{{Foster}} \textit{et~al.}},
\bibinfo{journal}{Phys. Rev. Lett.}
  \textbf{\bibinfo{volume}{85}}, \bibinfo{pages}{3149}
  (\bibinfo{year}{2000}{\natexlab{a}}).

\bibitem[{\citenamefont{{Foster}
  et~al.}(2000{\natexlab{b}})\citenamefont{{Foster}, {Mielke}, and
  {Orozco}}}]{Foster00a}
\bibinfo{author}{\bibfnamefont{G.~T.} \bibnamefont{{Foster}}},
  \bibinfo{author}{\bibfnamefont{S.~L.} \bibnamefont{{Mielke}}},
  \bibnamefont{and} \bibinfo{author}{\bibfnamefont{L.~A.}
  \bibnamefont{{Orozco}}}, \bibinfo{journal}{Phys. Rev. A}
  \textbf{\bibinfo{volume}{61}}, \bibinfo{pages}{053821}
  (\bibinfo{year}{2000}{\natexlab{b}}).

\bibitem[{\citenamefont{{Kuhn} et~al.}(2002)\citenamefont{{Kuhn}, {Hennrich},
  and {Rempe}}}]{Kuhn02}
\bibinfo{author}{\bibfnamefont{A.}~\bibnamefont{{Kuhn}}},
  \bibinfo{author}{\bibfnamefont{M.}~\bibnamefont{{Hennrich}}},
  \bibnamefont{and} \bibinfo{author}{\bibfnamefont{G.}~\bibnamefont{{Rempe}}},
  \bibinfo{journal}{Phys. Rev. Lett.} \textbf{\bibinfo{volume}{89}},
  \bibinfo{pages}{067901} (\bibinfo{year}{2002}).

\bibitem[{\citenamefont{{McKeever} et~al.}(2004)\citenamefont{{McKeever},
  {Boca}, {Boozer}, {Miller}, {Buck}, {Kuzmich}, and {Kimble}}}]{McKeever04}
\bibinfo{author}{\bibfnamefont{J.}~\bibnamefont{{McKeever}} \textit{et~al.}},
\bibinfo{journal}{Science}
  \textbf{\bibinfo{volume}{303}}, \bibinfo{pages}{1992} (\bibinfo{year}{2004}).

\bibitem[{\citenamefont{{Keller} et~al.}(2004)\citenamefont{{Keller}, {Lange},
  {Hayasaka}, {Lange}, and {Walther}}}]{Keller04}
\bibinfo{author}{\bibfnamefont{M.}~\bibnamefont{{Keller}} \textit{et~al.}},
  \bibinfo{journal}{Nature} \textbf{\bibinfo{volume}{431}},
  \bibinfo{pages}{1075} (\bibinfo{year}{2004}).

\bibitem[{\citenamefont{{Birnbaum} et~al.}(2005)\citenamefont{{Birnbaum},
  {Boca}, {Miller}, {Boozer}, {Northup}, and {Kimble}}}]{Birnbaum05}
\bibinfo{author}{\bibfnamefont{K.~M.} \bibnamefont{{Birnbaum}} \textit{et~al.}},
\bibinfo{journal}{Nature}
  \textbf{\bibinfo{volume}{436}}, \bibinfo{pages}{87} (\bibinfo{year}{2005}).

\bibitem[{\citenamefont{{Hijlkema} et~al.}(2007)\citenamefont{{Hijlkema},
  {Weber}, {Specht}, {Webster}, {Kuhn}, and {Rempe}}}]{Hijlkema07}
\bibinfo{author}{\bibfnamefont{M.}~\bibnamefont{{Hijlkema}} \textit{et~al.}},
  \bibinfo{journal}{Nature Phys.} \textbf{\bibinfo{volume}{3}},
  \bibinfo{pages}{253} (\bibinfo{year}{2007}).

\bibitem[{\citenamefont{Dayan et~al.}(2008)\citenamefont{Dayan, Parkins, Aoki,
  Ostby, Vahala, and Kimble}}]{Dayan08}
\bibinfo{author}{\bibfnamefont{B.}~\bibnamefont{Dayan} \textit{et~al.}},
\bibinfo{journal}{Science}
  \textbf{\bibinfo{volume}{319}}, \bibinfo{pages}{1062} (\bibinfo{year}{2008}).

\bibitem[{\citenamefont{{Chang} et~al.}(2007)\citenamefont{{Chang},
  {S{\o}rensen}, {Demler}, and {Lukin}}}]{Chang07}
\bibinfo{author}{\bibfnamefont{D.~E.} \bibnamefont{{Chang}} \textit{et~al.}},
\bibinfo{journal}{Nature Phys.}
  \textbf{\bibinfo{volume}{3}}, \bibinfo{pages}{807} (\bibinfo{year}{2007}).

\bibitem[{\citenamefont{{Schuster} et~al.}(2008)\citenamefont{{Schuster},
  {Kubanek}, {Fuhrmanek}, {Puppe}, {Pinkse}, {Murr}, and {Rempe}}}]{Schuster08}
\bibinfo{author}{\bibfnamefont{I.}~\bibnamefont{{Schuster}}},
  \bibinfo{author}{\bibfnamefont{A.}~\bibnamefont{{Kubanek}}},
  \bibinfo{author}{\bibfnamefont{A.}~\bibnamefont{{Fuhrmanek}}},
  \bibinfo{author}{\bibfnamefont{T.}~\bibnamefont{{Puppe}}},
  \bibinfo{author}{\bibfnamefont{P.~W.~H.} \bibnamefont{{Pinkse}}},
  \bibinfo{author}{\bibfnamefont{K.}~\bibnamefont{{Murr}}}, \bibnamefont{and}
  \bibinfo{author}{\bibfnamefont{G.}~\bibnamefont{{Rempe}}},
  \bibinfo{journal}{Nature Phys.} \textbf{\bibinfo{volume}{4}},
  \bibinfo{pages}{382} (\bibinfo{year}{2008}).

\bibitem[{\citenamefont{{Carmichael} et~al.}(1996)\citenamefont{{Carmichael},
  {Kochan}, and {Sanders}}}]{Carmichael96}
\bibinfo{author}{\bibfnamefont{H.~J.} \bibnamefont{{Carmichael}}},
  \bibinfo{author}{\bibfnamefont{P.}~\bibnamefont{{Kochan}}}, \bibnamefont{and}
  \bibinfo{author}{\bibfnamefont{B.~C.} \bibnamefont{{Sanders}}},
  \bibinfo{journal}{Phys. Rev. Lett.} \textbf{\bibinfo{volume}{77}},
  \bibinfo{pages}{631} (\bibinfo{year}{1996}).

\bibitem[{\citenamefont{Schneebeli et~al.}(2008)\citenamefont{Schneebeli, Kira,
  and Koch}}]{Schneebeli08}
\bibinfo{author}{\bibfnamefont{L.}~\bibnamefont{Schneebeli}},
  \bibinfo{author}{\bibfnamefont{M.}~\bibnamefont{Kira}}, \bibnamefont{and}
  \bibinfo{author}{\bibfnamefont{S.~W.} \bibnamefont{Koch}},
  \bibinfo{journal}{Phys. Rev. Lett.} \textbf{\bibinfo{volume}{101}},
  \bibinfo{pages}{097401} (\bibinfo{year}{2008}).

\bibitem[{\citenamefont{{Jaynes} and {Cummings}}(1963)}]{Jaynes63}
\bibinfo{author}{\bibfnamefont{E.~T.} \bibnamefont{{Jaynes}}} \bibnamefont{and}
  \bibinfo{author}{\bibfnamefont{F.~W.} \bibnamefont{{Cummings}}},
  \bibinfo{journal}{Proc. IEEE} \textbf{\bibinfo{volume}{51}},
  \bibinfo{pages}{89} (\bibinfo{year}{1963}).

\bibitem[{\citenamefont{{Mabuchi} and {Doherty}}(2002)}]{Mabuchi02}
\bibinfo{author}{\bibfnamefont{H.}~\bibnamefont{{Mabuchi}}} \bibnamefont{and}
  \bibinfo{author}{\bibfnamefont{A.~C.} \bibnamefont{{Doherty}}},
  \bibinfo{journal}{Science} \textbf{\bibinfo{volume}{298}},
  \bibinfo{pages}{1372} (\bibinfo{year}{2002}).

\bibitem[{\citenamefont{{Vahala}}(2003)}]{Vahala03}
\bibinfo{author}{\bibfnamefont{K.~J.} \bibnamefont{{Vahala}}},
  \bibinfo{journal}{Nature} \textbf{\bibinfo{volume}{424}},
  \bibinfo{pages}{839} (\bibinfo{year}{2003}).

\bibitem[{\citenamefont{{Khitrova} et~al.}(2006)\citenamefont{{Khitrova},
  {Gibbs}, {Kira}, {Koch}, and {Scherer}}}]{Khitrova06}
\bibinfo{author}{\bibfnamefont{G.}~\bibnamefont{{Khitrova}} \textit{et~al.}},
  \bibinfo{journal}{Nature Phys.} \textbf{\bibinfo{volume}{2}},
  \bibinfo{pages}{81} (\bibinfo{year}{2006}).

\bibitem[{\citenamefont{{Faraon} et~al.}(2008)\citenamefont{{Faraon},
  {Fushman}, {Englund}, {Stoltz}, {Petroff}, and {Vuckovic}}}]{Faraon08}
\bibinfo{author}{\bibfnamefont{A.}~\bibnamefont{{Faraon}} \textit{et~al.}},
\bibinfo{journal}{Nature Phys.}
(\bibinfo{year}{2008}), \eprint{doi:10.1038/nphys1078}.

\bibitem[{\citenamefont{{Rempe} et~al.}(1987)\citenamefont{{Rempe}, {Walther},
  and {Klein}}}]{Rempe87}
\bibinfo{author}{\bibfnamefont{G.}~\bibnamefont{{Rempe}}},
\bibinfo{author}{\bibfnamefont{H.} \bibnamefont{Walther}},
\bibinfo{author}{\bibfnamefont{N.} \bibnamefont{Klein}},
  \bibinfo{journal}{Phys. Rev. Lett.} \textbf{\bibinfo{volume}{58}},
  \bibinfo{pages}{353} (\bibinfo{year}{1987}).

\bibitem[{\citenamefont{Brune et~al.}(1996)\citenamefont{Brune, Schmidt-Kaler,
  Maali, Dreyer, Hagley, Raimond, and Haroche}}]{Brune96}
\bibinfo{author}{\bibfnamefont{M.}~\bibnamefont{Brune} \textit{et~al.}},
  \bibinfo{journal}{Phys. Rev. Lett.} \textbf{\bibinfo{volume}{76}},
  \bibinfo{pages}{1800} (\bibinfo{year}{1996}).

\bibitem[{\citenamefont{{Schuster} et~al.}(2007)\citenamefont{{Schuster},
  {Houck}, {Schreier}, {Wallraff}, {Gambetta}, {Blais}, {Frunzio}, {Majer},
  {Johnson}, {Devoret} et~al.}}]{Schuster07}
\bibinfo{author}{\bibfnamefont{D.~I.} \bibnamefont{{Schuster}} \textit{et~al.}},
\bibinfo{journal}{Nature}
  \textbf{\bibinfo{volume}{445}}, \bibinfo{pages}{515} (\bibinfo{year}{2007}).

\bibitem[{\citenamefont{Fink et~al.}(2008)\citenamefont{Fink, {G\"oppl}, Baur,
  Bianchetti, Leek, Blais, and Wallraff}}]{Fink08}
\bibinfo{author}{\bibfnamefont{J.~M.} \bibnamefont{Fink} \textit{et~al.} },
  \bibinfo{journal}{Nature} \textbf{\bibinfo{volume}{454}},
  \bibinfo{pages}{315} (\bibinfo{year}{2008}).

\bibitem[{\citenamefont{Hofheinz et~al.}(2008)\citenamefont{Hofheinz, Weig,
  Ansmann, Bialczak, Lucero, Neeley, O'Connell, Wang, Martinis, and
  Cleland}}]{Hofheinz08}
\bibinfo{author}{\bibfnamefont{M.}~\bibnamefont{Hofheinz} \textit{et~al.}},
\bibinfo{journal}{Nature}
  \textbf{\bibinfo{volume}{454}}, \bibinfo{pages}{310} (\bibinfo{year}{2008}).

\bibitem[{\citenamefont{{Carmichael} et~al.}(1991)\citenamefont{{Carmichael},
  {Brecha}, and {Rice}}}]{Carmichael91}
\bibinfo{author}{\bibfnamefont{H.~J.} \bibnamefont{{Carmichael}}},
  \bibinfo{author}{\bibfnamefont{R.~J.} \bibnamefont{{Brecha}}},
  \bibnamefont{and} \bibinfo{author}{\bibfnamefont{P.~R.}
  \bibnamefont{{Rice}}}, \bibinfo{journal}{Opt. Comm.}
  \textbf{\bibinfo{volume}{82}}, \bibinfo{pages}{73} (\bibinfo{year}{1991}).

\bibitem[{\citenamefont{Brecha et~al.}(1999)\citenamefont{Brecha, Rice, and
  Xiao}}]{Brecha99}
\bibinfo{author}{\bibfnamefont{R.~J.} \bibnamefont{Brecha}},
\bibinfo{author}{\bibfnamefont{P.~R.} \bibnamefont{Rice}},
\bibinfo{author}{\bibfnamefont{M.} \bibnamefont{Xiao}},
  \bibinfo{journal}{Phys. Rev. A} \textbf{\bibinfo{volume}{59}},
  \bibinfo{pages}{2392} (\bibinfo{year}{1999}).

\bibitem[{\citenamefont{{Goto} and {Ichimura}}(2004)}]{Goto04}
\bibinfo{author}{\bibfnamefont{H.}~\bibnamefont{{Goto}}} \bibnamefont{and}
  \bibinfo{author}{\bibfnamefont{K.}~\bibnamefont{{Ichimura}}},
  \bibinfo{journal}{\pra} \textbf{\bibinfo{volume}{70}},
  \bibinfo{pages}{023815} (\bibinfo{year}{2004}).

\bibitem[{\citenamefont{Puppe et~al.}(2007)\citenamefont{Puppe, Schuster,
  Grothe, Kubanek, Murr, Pinkse, and Rempe}}]{Puppe07}
\bibinfo{author}{\bibfnamefont{T.}~\bibnamefont{Puppe} \textit{et~al.}},
  \bibinfo{journal}{Phys. Rev. Lett.} \textbf{\bibinfo{volume}{99}},
  \bibinfo{eid}{013002} (\bibinfo{year}{2007}).

\bibitem[{\citenamefont{M{\"u}nstermann
  et~al.}(1999)\citenamefont{M{\"u}nstermann, Fischer, Maunz, Pinkse, and
  Rempe}}]{muenstermann99}
\bibinfo{author}{\bibfnamefont{P.}~\bibnamefont{M{\"u}nstermann} \textit{et~al.} },
  \bibinfo{journal}{Phys. Rev. Lett.} \textbf{\bibinfo{volume}{82}},
  \bibinfo{pages}{3791} (\bibinfo{year}{1999}).

\end{thebibliography}

\end{document}